\documentclass[%
 reprint,
 amsmath,amssymb,
 aps,prl,
]{revtex4-2}

\usepackage{graphicx}
\usepackage{dcolumn}
\usepackage{bm}
\usepackage{xcolor}
\usepackage{comment}
\usepackage{rotating}
\usepackage{multirow}

\newcommand{\aap}{Astron. Astroph.}

\begin{document}
\preprint{APS/123-QED}

\title{The role of the irreducible mass in repetitive Penrose energy extraction processes in a Kerr black hole} 

 \author{R. Ruffini$^{1,2,3,4}$}
 \email{ruffini@icra.it}
 \author{C. L. Bianco$^{1,2,3,5}$}
 \author{M. Prakapenia$^{1}$}
 \author{H. Quevedo$^{6,7,1}$}
 \author{J. A. Rueda$^{1,3,7,8,9}$}
\author{S. Zhang$^{1,8,9,10,11}$}

\affiliation{$^{1}$ ICRANet, Piazza della Repubblica 10, I-65122 Pescara, Italy}

\affiliation{$^{2}$ICRANet, 1 Avenue Ratti, 06000 Nice, France}

\affiliation{$^{3}$ICRA, Dipartimento di Fisica, Sapienza Universit$\grave{a}$ di Roma, Piazzale Aldo Moro 5, I-00185 Roma, Italy}

\affiliation{$^{4}$INAF, Viale del Parco Mellini 84, 00136 Rome, Italy}

\affiliation{$^{5}$INAF, Istituto di Astrofisica e Planetologia Spaziali, Via Fosso del Cavaliere 100, 00133 Rome, Italy}

\affiliation{$^{6}$Instituto de Ciencias Nucleares, Universidad Nacional Aut\'{o}noma de 
M\'{e}xico, AP 70543, Mexico City 04510, Mexico}

\affiliation{$^{7}$Al-Farabi Kazakh National University, Al-Farabi Ave. 71, 050040 Almaty, Kazakhstan}

\affiliation{$^{8}$ICRANet-Ferrara, Dipartimento di Fisica e Scienze della Terra, Universit\`a degli Studi di Ferrara, Via Saragat 1, I-44122 Ferrara, Italy}

\affiliation{$^{9}$Dipartimento di Fisica e Scienze della Terra, Universit\`a degli Studi di Ferrara, Via Saragat 1, I--44122 Ferrara, Italy}

\affiliation{$^{10}$School of Astronomy and Space Science, University of Science and Technology of China, Hefei 230026, China}

\affiliation{$^{11}$CAS Key Laboratory for Research in Galaxies and Cosmology, Department of Astronomy, University of Science and Technology of China, Hefei 230026, China}

\begin{abstract}
The concept of the irreducible mass ($M_{\rm irr}$) has led to the mass-energy ($M$) formula of a Kerr black hole (BH), in turn leading to its surface area $S=16\pi M_{\rm irr}^2$. This also allowed the coeval identification of the reversible and irreversible transformations, soon followed by the concepts of \textit{extracted} and \textit{extractable} energy. This new conceptual framework avoids inconsistencies recently evidenced in a repetitive Penrose process. We consider repetitive decays in the ergosphere of an initially extreme Kerr BH and show the processes are highly irreversible. For each decay, the particle that the BH captures causes an increase of the irreducible mass {(so the BH horizon)}, much larger than the extracted energy. The energy extraction process stops {when the BH reaches a positive spin lower limit set by the process boundary conditions}. Thus, the reaching of a final non-rotating Schwarzschild BH state through this accretion process is impossible. We have assessed such processes for selected decay radii {and incoming particle with rest mass $1 \%$ of the BH initial mass $M_0$. For $r= 1.2 M$ and $1.9 M$, the sequence stops after $8$ and $34$ decays, respectively, at a spin $0.991$ and $0.857$, the energy extracted has been only $1.16\%$, and $0.42\%$, the extractable energy is reduced by $17\%$ and $56\%$, and the irreducible mass increases by $5\%$ and $22\%$, all values in units of $M_0$. These results show the highly nonlinear change of the BH parameters, dictated by the BH mass-energy formula, and that the BH rotational energy is mainly converted into irreducible mass}. Thus, evaluating the irreducible mass increase in any energy extraction processes in the Kerr BH ergosphere is mandatory.
\end{abstract}

\date{\today}

\maketitle

\section{\label{sec:I}Introduction}

A recent article \cite{2024arXiv240508229R} (hereafter Paper I) has revisited the equations of a single Penrose process of a test particle $\mu_0$ that decays into two particles {of rest mass} $\mu_1$ and $\mu_2$, inside the ergosphere of an extreme Kerr black hole (BH) with mass $M$ and {angular momentum $L = M^2$ (i.e., with spin parameter $\hat a= a/M = L/M^2= 1$)}, adopting $\mu_0= 10^{-2} M$, to fulfill the condition of test particle. {Throughout, we use $c=G=1$ geometric units and denote with over hats dimensionless quantities. The four-momentum of the $i$-th particle is $p^\alpha_i = \mu_i u^\alpha_i$, where $u^\alpha_i$ is the four-velocity relative to an observer at rest at infinity. Thus, the particles' conserved energy and angular momentum are $E_i = -p_{t i}$ and $p_{\phi i}$, and their dimensionless values are $\hat E_i = E_i/\mu_i$, and $\hat p_{\phi i} = p_{\phi i}/(\mu_i M)$.} The decays at three radii inside the BH ergosphere were studied: $r=1.2 M$, $1.5 M$, and $1.9 M$. The basic assumption to describe the process was that the decay occurs at the turning points of their trajectories, and the energy of the decaying particle, $E_0 = \mu_0$, the angular momentum of the particle entering the BH, $p_{\phi1} = -19.434\,\mu_1 M$, and the mass ratio $\mu_2/\mu_1 = 0.78345$. This choice has the advantage of leading to a well-tested working solution of the decay equations. Paper I showed that if one turns from a single Penrose process, linear in the BH mass and angular momentum, {into a sequence of events (as suggested in some hypothetical Penrose process applications, e.g., \cite{Misner:1973prb}), keeping the above linearity} violates energy conservation. Thus, a nonlinear process treatment is necessary. {This paper aims to consider a sequence of Penrose processes to show the role of} the nonlinear relation between $M$, $a$, and the irreducible mass, $M_{\rm irr}$ \cite{1970PhRvL..25.1596C}. The latter classifies BH transformations into reversible ones, keeping $M_{\rm irr}$ constant, and irreversible ones where it increases.

To approach this problem, we express first the general equations for the mass-energy of a Kerr BH characterized by mass, angular momentum, and the irreducible mass \cite{1971PhRvD...4.3552C,1971PhRvL..26.1344H}, including the upper limits to the spin parameter and the expression of the horizon surface area. In addition to these classic results, we express the inverse relation relating the irreducible mass to the BH mass and spin. In total generality, we also introduce the extracted energies presented in Paper I and the novel concept of extractable energy. We recall how the maximum extractable energy cannot exceed $29.3\%$ of the BH mass of an extreme ($\hat a=1$) Kerr BH. Intermediate values apply for other BH parameters of the Kerr BH.

Next, we extend the analytic solution of the Penrose Process to any spin value. We show the appearance of the irreducible mass role in the first step, starting from an extreme Kerr BH, for the same fixed assumptions of Paper I. The fundamental result is that a Penrose process is far from linear; the correct theoretical approach shows its high nonlinearity and irreversibility; the irreducible mass increases much more than the extracted energy. 

We then show a repetitive nonlinear process, { formulating the BH changes in the iterative process from the $n-1$ to the $n$-th event. Table \ref{tab:Tablesequence12} summarizes the explicit example of a sequence of decays at $r=1.2 M$ and $r=1.9 M$. We then summarize the conclusions.} Due to the {substantial increase of the irreducible mass and decrease of the spin} at each step, the energy extraction {stops} after a finite number of steps. The final value of the extractable energy remains very close to the initial value, or it is reduced owing to the conversion of rotational energy into irreducible mass. In the former case, the remaining extractable energy can be extracted by a reversible electrodynamical process, as shown in a third article (Rueda and Ruffini, submitted). It remains to determine how the irreducible mass increase could affect any emission process occurring in the BH ergosphere.

\section{The irreducible mass}\label{sec:II}

{We start recalling the Kerr BH metric in Boyer-Lindquist coordinates, $ds^2 = g_{tt} dt^2 + g_{rr} dr^2 + g_{\theta\theta} d\theta^2 + g_{\phi\phi} d\phi^2 + 2 g_{t \phi} dt d\phi$, where $g_{tt} = -\left(1-2 M r/\Sigma\right)$, $g_{rr} = \Sigma/\Delta$, $g_{\theta\theta} = \Sigma$, $g_{\phi\phi} = A\sin^2\theta/\Sigma$, $g_{t \phi} = - 2 a M r \sin^2\theta/\Sigma$, being $\Sigma=r^2+a^2\cos^2\theta$, $\Delta=r^2-2 M r+ a^2$, and $A = (r^2+a^2)^2-\Delta a^2 \sin^2\theta$. The BH outer horizon (the largest root of $\Delta = 0$) is $r = r_+ = M+\sqrt{M^2-a^2}$.}  The mass-energy of the Kerr BH is \cite{1970PhRvL..25.1596C,1971PhRvD...4.3552C}
{
\begin{equation}\label{eq:M}
    M^2=M_{\rm irr}^2+\frac{L^2}{4 M_{\rm irr}^2},
\end{equation}
}
which expresses it as a function of the BH angular momentum $L$ and the irreducible mass, $M_{\rm irr}$. The latter must fulfill the condition {$L^2/(4 M_{\rm irr}^4) \leq 1$ (i.e., $\hat a^2 \leq 1$)} for stable configurations to exist. 

The irreducible mass holds constant in reversible transformations, while it monotonically increases in irreversible transformations. Soon after the introduction of the mass-energy formula (\ref{eq:M}), \citet{1971PhRvD...4.3552C} and \citet{1971PhRvL..26.1344H} introduced the relation between the irreducible mass and the surface area of the BH horizon $S = 16\pi M_{\rm irr}^2$, 
which for an infinitesimal change $\delta M_{\rm irr}$, increases by $\delta S = 32 \pi M_{\rm irr} \delta M_{\rm irr}$. Thus, it increases for irreversible transformations ($\delta M_{\rm irr} >0$) and constant for reversible ones ($\delta M_{\rm irr} =0$).

Inverting Eq. (\ref{eq:M}), {and selecting the physically-relevant root of the equation, i.e., that leading to $M_{\rm irr} = M$ for $\hat a = 0$, we obtain the expression of the irreducible mass in terms} of the BH mass and angular momentum 
{
\begin{equation}
     M_{\rm irr}=\sqrt {\frac{M r_+}{2}} = M \sqrt{\frac{1 + \sqrt{1-\hat a^2}}{2}},
    \label{eq:Mirr}
\end{equation}
where $r_+ = M \hat r_+ = M ( 1 + \sqrt{1-\hat a^2})$ is the Kerr BH outer event horizon.} With the irreducible mass, we introduce the concept of BH extractable energy {\cite{2019ApJ...886...82R,2021A&A...649A..75M}}
{
\begin{align}\label{eq:Eext}
    E_{\rm extractable}&\equiv  M -M_{\rm irr}= M\left(1 - \sqrt{\frac{1+\sqrt{1- \hat a^2}}{2}     
    }\right) 
    ,
\end{align}
}
which is the maximum energy that can be extracted from the Kerr BH. For {maximal rotation} $(\hat a = 1)$, the extractable energy is $E_{\rm extractable} = M(1-1/\sqrt{2}) \approx 0.293 M$.

\section{The first irreducible mass increase in the Penrose process}\label{sec:III}

We now specialize in the case of particle motion on the equatorial plane, {$\theta = \pi/2$}. In this case, the  system of equations of the Penrose process is given by {four-momentum conservation, which leads to the energy, angular and radial momentum conservation equations}
\begin{subequations}\label{eq:eq6}
    \begin{align}
\hat{E}_0&=\tilde{\mu}_1\hat{E}_1+\tilde{\mu}_2\hat{E}_2,\label{subeq:eq6b}\\
\hat{p}_{\phi0}&=\tilde{\mu}_1\hat{p}_{\phi1}+\tilde{\mu}_2\hat{p}_{\phi2},\label{subeq:eq6c}\\
    \hat{p}_{r0} &= \tilde{\mu}_1 \hat{p}_{r1} + \tilde{\mu}_2\hat{p}_{r2},
    \label{subeq:eq6d}
    \end{align}
\end{subequations}
where $\tilde \mu_1 = \mu_1/\mu_0$ and $\tilde \mu_2 = \mu_2/\mu_0$, {and $\hat p_{r i}=p_{r i}/\mu_i$ are the dimensionless particles' (covariant) radial momentum. The latter are obtained from the normalization condition $g_{\mu \nu} \hat p^\mu_i \hat p^\nu_i = -1$, which leads to
\begin{equation}\label{eq:pr}
    \hat p^2_{ri} = \frac{\hat r}{\hat \Delta^2} (\hat E_i - \hat V^+_{i})(\hat E_i - \hat V^-_{i}),\quad i=0,1,2,
\end{equation}
defining the effective potential of the radial motion \cite{landau2013classical,1971ESRSP..52...45R}
\begin{equation}\label{eq:Veff}
    \hat V^\pm_{i} = \hat \Omega\,\hat p_{\phi i} \pm  \frac{\hat r}{\hat A}\sqrt{\hat \Delta (\hat A+\hat r^2 \hat p_{\phi i}^2)},
\end{equation}
where $\hat \Delta = \Delta/M^2$, $\hat A = A/M^4$, and $\hat \Omega = -M\,g_{t\phi}/g_{\phi \phi} = 2 \hat a\,\hat r/\hat A$ is the spacetime angular velocity measured by the observer at rest at infinity.}

{
At this stage, assuming that we set a decay radius and the BH spin, the problem counts $11$ unknowns (two masses, three energies, three angular momenta, and three radial momenta) and only $6$ equations. Further, the equations must be solved along with boundary conditions imposed by the realization of the Penrose process, as follows. First, the mass defect must satisfy $\tilde \mu_{\rm def} = \mu_{\rm def}/\mu_0 = 1 - (\tilde \mu_1 + \tilde \mu_2) > 0$, which implies $\tilde \mu_1 + \tilde \mu_2 < 1$. Second, the condition for timelike geodesics reads $\gamma_i = (dt/d\tau)_i = \hat A(\hat E_i - \hat \Omega \hat p_{\phi i})/(\hat r^2 \hat \Delta) >0$, so it imposes $\hat E_i > \hat \Omega\,\hat p_{\phi i}$. For co-rotating and counter-rotating particles, $\hat V^+_{i} \geq \hat V^-_{i}$, so the condition $\hat p^2_{ri}\geq 0$ imposes $\hat E_i \geq \hat V^+_{i}$, which via Eq. (\ref{eq:Veff}), also implies $\gamma_i >0$. Now, energy must be extracted from the BH, so $E_2 > E_0$ and via Eq. (\ref{subeq:eq6b}), $E_1 < 0$, which occurs only inside the ergosphere, $\hat r_+ < \hat r < 2$. Therein, geodesics with $\hat E_1 < 0$ have $p_{\phi 1} < 0$. We are interested in the optimal conditions for maximum energy extraction. The conditions $\hat E_1 <0$ and $\hat E_1 \geq \hat V^+_1$ imply that $|\hat E_1|$ is maximal when particle ``1'' is at a turning point, i.e., $\hat p_{r1} = 0$. In this case, the radial momentum conservation (\ref{subeq:eq6d}) implies $p_{r0} = p_{r2}$. If the momenta were positive or negative, the above equality implies, via Eq. (\ref{eq:pr}), the two equations $E_0 - V_0^{\pm} = E_2 - V_2^{\pm}$ which by energy and angular momentum conservation, Eqs. (\ref{subeq:eq6b}) and (\ref{subeq:eq6c}), implies $\hat E_1 = \hat \Omega \hat p_{\phi1}$ that violates the timelike geodesics condition $\gamma_1 > 0$. Hence, $p_{r0} = p_{r2} = 0$ at the split point, and we conclude that all particles are at their turning points, i.e., $\hat E_i = \hat V^+_i$, for $i=0$, $1$, and $2$.
}

{
The above choice leaves us with $8$ unknowns and $5$ equations. We must assign values to three variables and solve the equations for the other five. Without loss of generality, we shall assume values for $\hat E_0$, $\hat p_{\phi1}$, and $\nu\equiv \mu_2/\mu_1$. The system of equations has the analytic solution
}
{
\begin{subequations}\label{eqs:solution}
    \begin{align}
        \hat p_{\phi0} &= \frac{2 \hat a {\hat E_0}-\sqrt{\hat r \hat \Delta [2 + \hat r (\hat E_0^2-1)]}}{2-\hat r},\\
        \hat E_1 &= \hat V_1^+ = \hat \Omega\,\hat p_{\phi 1} +  \frac{\hat r}{\hat A}\sqrt{\hat \Delta (\hat A+\hat r^2 \hat p_{\phi 1}^2)} ,\label{eqs:solutionE1}\\
       \tilde{\mu}_1 &={\frac{\hat r \hat \Delta}{K + \sqrt{K^2-\hat r^2 \hat\Delta^2 (1-\nu^2)}},}\\
        \hat E_2 &= {\frac{\hat E_0}{\tilde\mu_2}-\frac{\hat E_1}{\nu}},\quad 
        \hat p_{\phi2} = {\frac{\hat p_{\phi0}}{\tilde\mu_2}-\frac{\hat p_{\phi1}}{\nu}},
    \end{align}
\end{subequations}
}
with $K = \hat A \,\hat E_0 \hat E_1/\hat r - 2 \hat a (\hat E_0 \hat p_{\phi1} + \hat E_1 \hat p_{\phi0}) + (2-\hat r) \hat p_{\phi0}\,\hat p_{\phi1}$.

We denote with $M_0$ and $L_0$ the initial (i.e., before the process) mass and angular momentum, so $\hat a_0 = L_0/M_0^2$, $r_{+,0} = M_0 (1 + \sqrt{1-\hat a_0^2})$, and $M_{\rm irr,0} = M_0 \sqrt{\hat r_{+,0}/2}$. After capturing particle ``1'', the BH has the new parameters
\begin{subequations}\label{eqs:ML}
    \begin{align}
        M &= M_0 + \Delta M_0,\quad  \Delta M_0 = \hat E_{1} \mu_{1},\\
        L &= L_0 + \Delta L_0,\quad  \Delta L_0 = \hat p_{\phi1} \mu_{1} M_0.
    \end{align}
\end{subequations}
For the Penrose process, $\hat E_1 < 0$ and $\hat p_{\phi1}<0$, so $\Delta M_0 < 0$ and $\Delta L_0 < 0$, hence reducing the BH mass and angular momentum, $M < M_0$ and $L < L_0$. Accordingly, we can compute the changes in the spin parameter, event horizon, and irreducible mass change by
\begin{subequations}\label{eqs:arhorMirr}
    \begin{align}
        \Delta \hat a_0 & = \frac{L}{M^2} - \frac{L_0}{M_0^2},\\
        \Delta r_{+,0} &= M (1+\sqrt{1-\hat a^2}) - M_0 (1+\sqrt{1-\hat a_0^2}),\\
        \Delta M_{{\rm irr},0} &= M \sqrt{\frac{1+\sqrt{1-\hat a^2}}{2}} - M_0 \sqrt{\frac{1+\sqrt{1-\hat a_0^2}}{2}},
    \end{align}
\end{subequations}
With the above, we can also define the change in the extractable energy as
\begin{equation}\label{eq:DeltaEext}
    \Delta E_{{\rm extractable},0} = \Delta M_0 - \Delta M_{{\rm irr},_0},
\end{equation}
while the actual \textit{extracted energy} in the process is
\begin{equation}\label{eq:Eextracted}
    E_{{\rm extracted},0} = |\Delta M_0| = -\Delta M_0 = -E_1 = -\hat E_1 \mu_1.
\end{equation}

For example, let us start with an extreme Kerr BH, i.e., $\hat a_0 = 1$. The BH horizon is $r_{+,0} = M_0$ ($\hat r_{+,0} = 1$) and the irreducible mass $M_{{\rm irr},0} = M_0/\sqrt{2}$. The horizon surface area is $S_0=16\pi M_{{\rm irr,0}}^2 = 8\pi M_0^2 = 25.133 M_0^2$, and the initial  extractable energy is $E_{{\rm extractable},0} = M_0 - M_{{\rm irr,0}} = (1-\sqrt{2}/2)M_0 = 0.293 M_0$ (cf. Eq. \ref{eq:Eext}).

Assuming $\hat p_{\phi1} =-19.434$, $\nu=\mu_2/\mu_1 = 0.78345$, we obtain from Eqs. (\ref{eqs:solution}) the rest of parameters, {for given values of $\hat E_0$ and $\hat r$}. Table \ref{tab:table1} summarizes the results for $\hat r = 1.2$, $1.5$, and $1.9$, for $\hat E_0 = 1$ (at rest at infinity) and $1.5$ (launched with nonzero kinetic energy). {To quantify the change in the BH parameters, we set the decaying particle mass, $\mu_0 = 10^{-2} M_0$, in all cases. Notice all solutions of Table \ref{tab:table1} satisfy the mass-defect constraint $\tilde\mu_{\rm def}>0$, which translates into $\tilde\mu_1 < 1/(1+\mu_2/\mu_1) = 0.56071$. Further, our numerical tests show that different $\hat p_{\phi 1}$ and $\mu_2/\mu_1$ lead to different values of $\hat E_1$, $\hat E_2$, $\hat p_{\phi 2}$, $\mu_1$ and $\mu_2$, but give the same value of the physical energies $E_1 = \mu_1 \hat E_1$, $E_2 = \mu_2 \hat E_2$ and angular momenta, $p_{\phi 1} = \mu_1 M \hat p_{\phi 1}$, and $p_{\phi 2} = \mu_2 M \hat p_{\phi 2}$.}

In the case $\hat r = 1.2$ and $\hat E_0 = 1$, the extracted energy, $E_{{\rm extracted},0} = |\Delta M_0| = 0.00145 M_0$, is much lower than the increase in the irreducible mass, $\Delta M_{{\rm irr},0} = 0.0158 M_0$, which has increased more than ten times. {Non-zero initial velocities of particle ``0'' extract even less BH energy (see case $\hat E_0 = 1.5$ in Table \ref{tab:table1}). The above results} demonstrate the process inefficiency, which converts the BH rotational energy mostly into irreducible mass. The process is highly irreversible. We shall return to this point below. 

\begin{table*}
\caption{\label{tab:table1}%
Summary of Penrose process parameters occurring inside the ergosphere of extreme Kerr BH of mass $M_0$ and angular momentum $L_0=M_0^2$. The BH captures the decay-product particle with energy $E_1=\hat E_1 \mu_1 <0$ and angular momentum $p_{\phi1}=\hat p_{\phi 1}\mu_1 M_0<0$. We list the results for the decay occurring at three selected positions: $\hat{r}=1.2$, $1.5$, and $1.9$. We choose $\hat p_{\phi1} =-19.434$, $\nu = \mu_2/\mu_1 = 0.78345$, {$\hat E_0 = 1.0$ (upper) and $1.5$ (lower) and obtain the other parameters from the conservation equations and the BH parameters from Eqs. (\ref{eqs:solution})--(\ref{eq:Eextracted})}. We adopt $\mu_0 = 10^{-2}M_0$ for all cases.}
\begin{ruledtabular}
{
\begin{tabular}{c|ccccccc|ccccc}
$\hat r$ 
& $\displaystyle\frac{\mu_1}{\mu_0}$ 
& $\displaystyle\frac{\mu_2}{\mu_0}$
& $\displaystyle\frac{E_1}{\mu_0}$ 
& $\displaystyle\frac{E_2}{\mu_0}$ 
& $\displaystyle\frac{p_{\phi0}}{\mu_0}$ 
& $\displaystyle\frac{p_{\phi1}}{\mu_0}$ 
& $\displaystyle\frac{p_{\phi2}}{\mu_0}$ 
& $\displaystyle\frac{M}{M_0}$ 
& $\hat a$
& $\displaystyle\frac{\Delta M_{{\rm irr},0}}{M_0}$ 
& $\displaystyle\frac{E_{{\rm extracted},0}}{M_0}$ 
& $\displaystyle\frac{E_{\rm extractable}}{M_0}$\\
[6pt]
\hline
\multicolumn{13}{c}{$\hat E_0 = 1$}\\
\hline
$1.2$ & $0.02090$ & $0.01637$ & $-0.14496$ & $1.14496$ & $2.11270$ & $-0.40618$ & $2.51888$ & $0.99855$ & $0.99883$ & $0.01583$ & $0.00145$ & $0.27561$ \\
$1.5$ & $0.02182$ & $0.01709$ &$-0.07681$ & $1.07681$ & $2.26795$ & $-0.42402$ & $2.69197$ & $0.99923$ & $0.99729$ & $0.02498$ & $0.00077$ & $0.26714$ \\
$1.9$ & $0.02470$ & $0.01935$ & $-0.01237$ & $1.01237$ & $2.45577$ & $-0.47993$ & $2.93570$ & $0.99988$ & $0.99545$ & $0.03284$ & $0.00012$ & $0.25993$ \\
\hline
\multicolumn{13}{c}{$\hat E_0 = 1.5$}\\
\hline
$1.2$ & $0.01493$ & $0.01170$ & $-0.10354$ & $1.60354$ & $3.23765$ & $-0.29013$ & $3.52778$ & $0.99896$ & $0.99917$ & $0.01354$ & $0.00104$ & $0.27832$ \\
$1.5$ & $0.01513$ & $0.01186$ & $-0.05327$ & $1.55327$ & $3.58909$ & $-0.29408$ & $3.88316$ & $0.99947$ & $0.99812$ & $0.02095$ & $0.00053$ & $0.27142$ \\
$1.9$ & $0.01658$ & $0.01299$ & $-0.00830$ & $1.50830$ & $4.05173$ & $-0.32230$ & $4.37403$ & $0.99992$ & $0.99694$ & $0.02705$ & $0.00008$ & $0.26576$
\end{tabular}
}
\end{ruledtabular}
\end{table*}

\section{The repetitive nonlinear process}
\label{sec:IV}

Paper I studied the case of repetitive decays, according to which it would be possible to extract the entire rotational energy of an extreme Kerr BH. The process consisted of dropping, one by one, a particle ``0'' of energy $E_0 = \mu_0 = 10^{-2}M_0$, which decays inside the ergosphere into two particles ``1'' and ``2'' with energies $E_1 <0$ and $E_2>0$. All decays occur at the same location, $\hat r$. For $\hat r = 1.2$, it was shown the BH reaches the state of a non-rotating Schwarzschild BH after $246$ decays, assuming the repetitive process is linear in the mass and the angular momentum. However, Paper I showed that such a linear procedure violates energy conservation. This section shows the correct way to implement consecutive decays: {at each step, the BH mass and angular momentum must be evaluated, introducing a nonlinear dependence that coincides with assuming the BH mass-energy formula.}

We start from a BH with $M_0$ and $L_0$, and the same fixed particle parameters as above; $\hat E_0 = 1$, $\hat p_{\phi1} =-19.434$, $\nu = 0.78345$. The other parameters are given by the solution (\ref{eqs:solution}). The first decay was described in the previous section. We can now extend Eqs. (\ref{eqs:ML}) and (\ref{eqs:arhorMirr}) to a sequence of decays. The BH mass and angular momentum at the $n$th process depends on the energy $E_{1,n-1}$ and angular momentum $p_{\phi1}$ of the particle that the BH captured in the preceding $n-1$ process, as follows
\begin{subequations}
    \begin{align}
        M_n &= M_{n-1} + \Delta M_{n-1},\quad \Delta M_{n-1} = \hat E_{1,n-1} \mu_{1,n-1},\\
        L_n &= L_{n-1} + \Delta L_{n-1},\quad \Delta L_{n-1} = \hat p_{\phi1} \mu_{1,n-1} M_{n-1},
    \end{align}
\end{subequations}
leading to the changes in $\hat a$, $\hat r_+$, and $M_{\rm irr}$
\begin{subequations}
    \begin{align}
        \Delta \hat a_{n-1} &= \frac{L_n}{M_n^2} - \frac{L_{n-1}}{M_{n-1}^2},\\
        \Delta r_{+,n-1} &= M_n \left(1+\sqrt{1-\hat a_n^2}\right) - M_0 \left(1+\sqrt{1-\hat a_0^2}\right),\\
        \Delta M_{{\rm irr}, n-1} &= M_n \sqrt{\frac{1+\sqrt{1-\hat a_n^2}}{2}} - M_0 \sqrt{\frac{1+\sqrt{1-\hat a_0^2}}{2}}.
    \end{align}
\end{subequations}

{Table \ref{tab:Tablesequence12}} shows the results of the above procedure for a processes sequence at $r_n = 1.2 M_n$ {and $r_n = 1.9 M_n$. These examples show how accounting for} $M_n$ and $\hat a_n$ change in each process implies the nonlinear increase of $M_{{\rm irr},n}$, consistently with the BH mass-energy formula and the BH surface area theorem. {Further, the process cannot be continued arbitrarily. The reduction of $\hat a$ changes the effective potentials $\hat V_i^+$, leading to a positive lower limit to the spin, as follows. Particles ``0'' and ``2'' turning points must be on the right side of the effective potential maximum, $V^+_{i,{\rm max}}$. For the case $\hat E_0 = 1$, the above implies a spin lower limit $\hat a_{\rm min,0} = 2\sqrt{\hat r} -\hat r$, approached when $V^+_{0,{\rm max}} = \hat E_0 = 1$. If one allows $\hat E_0 >1$, the limit situation $V^+_{2,{\rm max}} = \hat E_2$ imposes the spin lower limit $\hat a_{\rm min, 2} = (3 - \hat r) \sqrt{\hat r}/2$. Particle ``1'' turning point exists if the discriminant of the square root of Eq. (\ref{eqs:solutionE1}) is positive, implying $\hat a_{\rm min,1} = \sqrt{\hat r (2-\hat r)}$, which occurs when $\Delta = 0$ is approached, i.e., when $\hat r_+ = \hat r$. It is easy to check that $\hat a_{\rm min,0} > \hat a_{\rm min,2} > \hat a_{\rm min,1}$ for $1 <\hat r <2$. Thus, the relevant spin limit in the present situation is $\hat a \geq \hat a_{\rm min,0}$. For lower spins, the solution (\ref{eqs:solution}) leads to the turning point of particle ``0'' to occur on the left-side of $\hat V_0^+$ (see, e.g., Supplementary Material \cite{supp}).}

\begin{figure}
    \centering
    \includegraphics[width=\hsize,clip]{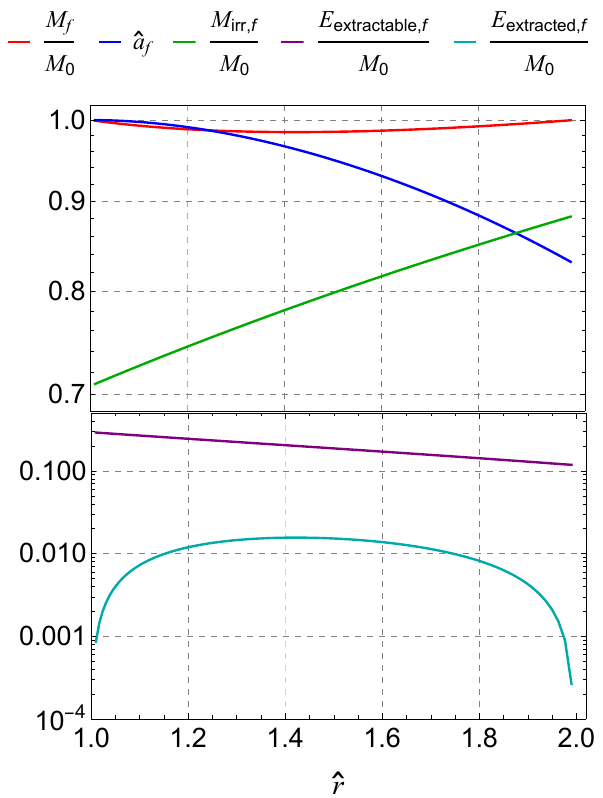}
    \caption{BH properties at {the end of the Penrose process sequence, for $\hat E_0 = 1$, $\hat p_{\phi 1}=-19.434$, $\mu_2/\mu_1 = 0.78345$, and $\mu_0/M_0 = 10^{-2}$. The decay radii vary from $\hat r = 1.01$ to $1.99$. Upper: BH mass, spin, and irreducible mass. Lower: Extractable and extracted energy. Notice the different scales of the vertical axis of the upper and lower plots}.}
    \label{fig:finalvalues}
\end{figure}

{
The process sequence is discrete, so $\hat a_{\rm min,0}$ is reached from above at a step $n_f$, where $\hat a_f = a_n (n = n_f) \gtrsim \hat a_{\rm min,0}$. A further process $n_f+1$ would lead to $\hat a < \hat a_{\rm min,0}$. Thus, the final BH parameters at the end of the sequence of processes are $M_f = M_n(n = n_f)$ and $L_f = L_n(n = n_f)$. In the example of decays at $\hat r_n = 1.2$, $\hat a_f = 0.9910$ is reached at $n_f=8$, and $\hat a_{\rm min,0} = 0.9909$. The decay sequence at $\hat r = 1.9$ ends after $n_f = 34$ steps with $\hat a_f = 0.8573$ ($\hat a_{\rm min,0} = 0.8568$). We could continue the sequence by 1) selecting a larger decay radius or 2) launching the splitting particle with $E_0>1$. The features of the former situation can be inferred from Fig. \ref{fig:finalvalues}, and the latter is less efficient, as shown in Table \ref{tab:table1}.
}

Figure \ref{fig:finalvalues} shows the final BH parameters after the sequence of Penrose processes, following the analogous calculation described above, for decay radii {in the range $\hat r = 1.01$ to $1.99$}. The plot shows that the larger the decay radius, the lower the final spin, so the lower the rotational and extractable energy. {The larger the radius, the lower the final spin at the end of the sequence, so} one could think the process extracts the rotational energy efficiently. However, inspecting the irreducible mass shows it becomes larger and larger, approaching the BH mass and explaining the decrease of the extractable energy. {Notice that also the extracted energy decreases for larger radii.} The above demonstrates a dramatic conclusion for the Penrose process: instead of being transferred to the escaping particle, the BH rotational energy lost mostly converts into irreducible mass.

\section{Conclusions}\label{sec:V}

We have considered self-consistently the BH mass and angular momentum changes at every step in a repetitive sequence of Penrose processes in the ergosphere of an initially extreme Kerr BH. We have shown that this leads to a nonlinear increase of the BH irreducible mass (so the horizon size), consistent with the BH mass-energy formula. The decrease of the BH spin leads to the end of the repetitive process {when it reaches a lower limit below which the process boundary conditions are not satisfied, leading} to significant consequences.
\begin{enumerate}
    \item The total energy extracted is only a tiny fraction of the initial Kerr BH mass. 
    \item For decays near the BH horizon (e.g., $\hat r=1.2$), the {spin lower limit is reached after a few decays. It leads to low energy extracted ($\lesssim 1\%$)}, so the BH parameters remain close to their initial values.
    \item The larger the decay radius {(e.g., $\hat r=1.9$)}, the larger the number of processes until the sequence stops. The final spin and the extractable energy reduce, and the extracted energy remains tiny ($\lesssim 0.4\%$), so the BH rotational energy converts into irreducible mass (see Fig. \ref{fig:finalvalues}). 
\end{enumerate}

Thus, the repetitive Penrose processes either produce a tiny change of the BH spin and rotational energy, leaving most of it to be extracted (by other means), or {it} converts the BH rotational energy mostly into irreducible mass.

We conclude that a necessary condition to analyze any {energy extraction process from a Kerr BH (e.g., \cite{piran1975high,2009PhRvL.103k1102B,2014PhRvL.113z1102S}), has necessary to account for the BH feedback via} the BH mass-energy formula, considering the nonlinear relation between $M$, $a$, and $M_{\rm irr}$. Consequently, we are brought to the quest for alternative BH energy extraction processes that keep the irreducible mass increase as small as possible. Electrodynamical processes in the vicinity of a Kerr BH embedded in a magnetic field and ionized matter offer a promising alternative: {in the sequence of Penrose process shown here, the increase of the BH irreducible mass relative to the decrease of the mass reaches values as large as $\Delta M_{\rm irr}/|\Delta M| \sim 500\%$, while in the electrodynamical process shown in \cite{2023EPJC...83..960R}, $\Delta M_{\rm irr}/|\Delta M| \sim 1$--$10\%$ (see, also, Rueda and Ruffini, submitted)}.

\begin{acknowledgments}
{We thank both referees for the careful analysis, reports, and relevant suggestions.}
We thank Prof. D. Christodoulou and Prof. R. P. Kerr for discussions on the topic of this work. The work of HQ was supported by UNAM PASPA-DGAPA.
\end{acknowledgments}

\clearpage
\begin{turnpage}
\begin{table}
\caption{\label{tab:Tablesequence12}%
Parameters for a repetitive Penrose process at $\hat r_n=1.2$ {and $1.9$. We set s fixed parameters,} $\hat E_0 = 1$, $\hat p_{\phi 1}= -19.344$ and $\mu_2/\mu_1 = 0.78345$. {To quantify the BH parameters, we assign to} the incoming particle mass, $\mu_0 = 10^{-2}M_0$. The remaining parameters {are obtained from the solution of equations governing the Penrose process at each step for the new BH mass and spin. The process efficiency parameter $\eta_n \equiv |E_1|/E_0 = -E_1/\mu_0$, so to get its percentage is sufficient to multiply the absolute value of the $14$th column by $100\%$}. 
}
\centering
{\scriptsize
\begin{ruledtabular}
{
\begin{tabular}{rrrrrrrrrrrrrrr}
$n$ &
$\displaystyle\frac{M_n}{M_0}$ & 
$\displaystyle\hat a_n$ & 
$\displaystyle\frac{E_{\rm extractable},n}{M_0} $    & 
$\displaystyle\frac{E_{\rm extracted},n}{M_0}$  & 
$\displaystyle\frac{M_{{\rm irr},n}}{M_0}$  & 
$\displaystyle\frac{r_{+,n}}{M_0}$  &
$\displaystyle\frac{\mu_{1,n}}{\mu_0}$ & 
$\displaystyle\frac{\mu_{2,n}}{\mu_0}$ &
$\displaystyle\frac{p_{\phi0,n}}{\mu_0}$ & 
$\displaystyle\frac{p_{\phi1,n}}{\mu_0}$  & 
$\displaystyle\frac{p_{\phi2,n}}{\mu_0}$  & 
$\displaystyle\frac{E_{0,n}}{\mu_0}$ & 
$\displaystyle\frac{E_{1,n}}{\mu_0}$  & 
$\displaystyle\frac{E_{2,n}}{\mu_0}$\\
 [6pt]
\hline
\multicolumn{15}{c}{$\hat r = 1.2$}\\
\hline
     0 & 1.000000 & 1.000000 & 0.292893 & 0.000000 & 0.707107 & 1.000000 & 0.020901 & 0.016375 & 2.112702 & -0.406182 & 2.518883 & 1.000000 & -0.144958 & 1.144958 \\
 1 & 0.998550 & 0.998832 & 0.275611 & 0.001450 & 0.722940 & 1.046802 & 0.020813 & 0.016306 & 2.121255 & -0.404471 & 2.525727 & 1.000000 & -0.144962 & 1.144962 \\
 2 & 0.997101 & 0.997676 & 0.268419 & 0.002899 & 0.728682 & 1.065042 & 0.020723 & 0.016236 & 2.130062 & -0.402740 & 2.532802 & 1.000000 & -0.144967 & 1.144967 \\
 3 & 0.995651 & 0.996532 & 0.262915 & 0.004349 & 0.732736 & 1.078496 & 0.020633 & 0.016165 & 2.139141 & -0.400987 & 2.540128 & 1.000000 & -0.144971 & 1.144971 \\
 4 & 0.994201 & 0.995402 & 0.258294 & 0.005799 & 0.735907 & 1.089436 & 0.020542 & 0.016093 & 2.148517 & -0.399208 & 2.547725 & 1.000000 & -0.144976 & 1.144976 \\
 5 & 0.992752 & 0.994284 & 0.254245 & 0.007248 & 0.738506 & 1.098747 & 0.020449 & 0.016021 & 2.158217 & -0.397401 & 2.555618 & 1.000000 & -0.144981 & 1.144981 \\
 6 & 0.991302 & 0.993180 & 0.250608 & 0.008698 & 0.740694 & 1.106883 & 0.020354 & 0.015946 & 2.168273 & -0.395563 & 2.563836 & 1.000000 & -0.144986 & 1.144986 \\
 7 & 0.989852 & 0.992089 & 0.247286 & 0.010148 & 0.742566 & 1.114114 & 0.020258 & 0.015871 & 2.178722 & -0.393689 & 2.572411 & 1.000000 & -0.144990 & 1.144990 \\
 8 & 0.988402 & 0.991013 & 0.244218 & 0.011598 & 0.744184 & 1.120617 & 0.020159 & 0.015794 & 2.189609 & -0.391776 & 2.581385 & 1.000000 & -0.144995 & 1.144995\\ 
\hline
\multicolumn{15}{c}{$\hat r = 1.9$}\\
\hline
  0 & 1.000000 & 1.000000 & 0.292893 & 0.000000 & 0.707107 & 1.000000 & 0.024695 & 0.019347 & 2.455770 & -0.479926 & 2.935696 & 1.000000 & -0.012366 & 1.012366 \\
 1 & 0.999876 & 0.995447 & 0.259928 & 0.000124 & 0.739948 & 1.095182 & 0.020813 & 0.016306 & 2.121255 & -0.404471 & 2.525727 & 1.000000 & -0.144962 & 1.144962 \\
 2 & 0.999753 & 0.990916 & 0.246786 & 0.000247 & 0.752967 & 1.134198 & 0.020723 & 0.016236 & 2.130062 & -0.402740 & 2.532802 & 1.000000 & -0.144967 & 1.144967 \\
 3 & 0.999629 & 0.986408 & 0.236921 & 0.000371 & 0.762708 & 1.163880 & 0.020633 & 0.016165 & 2.139141 & -0.400987 & 2.540128 & 1.000000 & -0.144971 & 1.144971 \\
 4 & 0.999505 & 0.981923 & 0.228757 & 0.000495 & 0.770748 & 1.188693 & 0.020542 & 0.016093 & 2.148517 & -0.399208 & 2.547725 & 1.000000 & -0.144976 & 1.144976 \\
 5 & 0.999381 & 0.977460 & 0.221684 & 0.000619 & 0.777697 & 1.210373 & 0.020449 & 0.016021 & 2.158217 & -0.397401 & 2.555618 & 1.000000 & -0.144981 & 1.144981 \\
 6 & 0.999257 & 0.973018 & 0.215389 & 0.000743 & 0.783868 & 1.229813 & 0.020354 & 0.015946 & 2.168273 & -0.395563 & 2.563836 & 1.000000 & -0.144986 & 1.144986 \\
 7 & 0.999133 & 0.968599 & 0.209683 & 0.000867 & 0.789450 & 1.247544 & 0.020258 & 0.015871 & 2.178722 & -0.393689 & 2.572411 & 1.000000 & -0.144990 & 1.144990 \\
 8 & 0.999009 & 0.964202 & 0.204446 & 0.000991 & 0.794563 & 1.263913 & 0.020159 & 0.015794 & 2.189609 & -0.391776 & 2.581385 & 1.000000 & -0.144995 & 1.144995 \\
 9 & 0.998885 & 0.959827 & 0.199593 & 0.001115 & 0.799292 & 1.279163 & 0.023596 & 0.018487 & 2.526752 & -0.458570 & 2.985323 & 1.000000 & -0.012421 & 1.012421 \\
 10 & 0.998761 & 0.955474 & 0.195061 & 0.001239 & 0.803700 & 1.293470 & 0.023477 & 0.018393 & 2.534993 & -0.456249 & 2.991242 & 1.000000 & -0.012426 & 1.012426 \\
 11 & 0.998636 & 0.951143 & 0.190804 & 0.001364 & 0.807832 & 1.306967 & 0.023358 & 0.018300 & 2.543305 & -0.453937 & 2.997243 & 1.000000 & -0.012432 & 1.012432 \\
 12 & 0.998512 & 0.946833 & 0.186787 & 0.001488 & 0.811725 & 1.319759 & 0.023239 & 0.018207 & 2.551690 & -0.451636 & 3.003326 & 1.000000 & -0.012438 & 1.012438 \\
 13 & 0.998388 & 0.942545 & 0.182979 & 0.001612 & 0.815408 & 1.331929 & 0.023122 & 0.018115 & 2.560149 & -0.449344 & 3.009493 & 1.000000 & -0.012443 & 1.012443 \\
 14 & 0.998263 & 0.938278 & 0.179358 & 0.001737 & 0.818905 & 1.343545 & 0.023004 & 0.018023 & 2.568680 & -0.447063 & 3.015743 & 1.000000 & -0.012449 & 1.012449 \\
 15 & 0.998139 & 0.934032 & 0.175904 & 0.001861 & 0.822235 & 1.354663 & 0.022887 & 0.017931 & 2.577286 & -0.444791 & 3.022077 & 1.000000 & -0.012454 & 1.012454 \\
 16 & 0.998014 & 0.929808 & 0.172600 & 0.001986 & 0.825414 & 1.365328 & 0.022771 & 0.017840 & 2.585966 & -0.442529 & 3.028495 & 1.000000 & -0.012460 & 1.012460 \\
 17 & 0.997890 & 0.925605 & 0.169434 & 0.002110 & 0.828456 & 1.375582 & 0.022655 & 0.017749 & 2.594721 & -0.440277 & 3.034998 & 1.000000 & -0.012465 & 1.012465 \\
 18 & 0.997765 & 0.921423 & 0.166392 & 0.002235 & 0.831373 & 1.385457 & 0.022540 & 0.017659 & 2.603551 & -0.438035 & 3.041586 & 1.000000 & -0.012470 & 1.012470 \\
 19 & 0.997640 & 0.917262 & 0.163466 & 0.002360 & 0.834174 & 1.394984 & 0.022425 & 0.017569 & 2.612457 & -0.435802 & 3.048259 & 1.000000 & -0.012476 & 1.012476 \\
 20 & 0.997515 & 0.913122 & 0.160646 & 0.002485 & 0.836869 & 1.404188 & 0.022310 & 0.017479 & 2.621439 & -0.433579 & 3.055018 & 1.000000 & -0.012481 & 1.012481 \\
 21 & 0.997391 & 0.909003 & 0.157925 & 0.002609 & 0.839466 & 1.413092 & 0.022196 & 0.017390 & 2.630498 & -0.431365 & 3.061864 & 1.000000 & -0.012486 & 1.012486 \\
 22 & 0.997266 & 0.904905 & 0.155295 & 0.002734 & 0.841971 & 1.421716 & 0.022083 & 0.017301 & 2.639635 & -0.429161 & 3.068796 & 1.000000 & -0.012491 & 1.012491 \\
 23 & 0.997141 & 0.900827 & 0.152751 & 0.002859 & 0.844390 & 1.430078 & 0.021970 & 0.017212 & 2.648849 & -0.426966 & 3.075815 & 1.000000 & -0.012496 & 1.012496 \\
 24 & 0.997016 & 0.896770 & 0.150287 & 0.002984 & 0.846729 & 1.438193 & 0.021858 & 0.017124 & 2.658141 & -0.424781 & 3.082922 & 1.000000 & -0.012501 & 1.012501 \\
 25 & 0.996891 & 0.892733 & 0.147898 & 0.003109 & 0.848993 & 1.446076 & 0.021746 & 0.017037 & 2.667512 & -0.422605 & 3.090118 & 1.000000 & -0.012506 & 1.012506 \\
 26 & 0.996766 & 0.888717 & 0.145579 & 0.003234 & 0.851187 & 1.453739 & 0.021634 & 0.016949 & 2.676963 & -0.420439 & 3.097402 & 1.000000 & -0.012511 & 1.012511 \\
 27 & 0.996641 & 0.884721 & 0.143327 & 0.003359 & 0.853313 & 1.461196 & 0.021523 & 0.016862 & 2.686493 & -0.418282 & 3.104775 & 1.000000 & -0.012516 & 1.012516 \\
 28 & 0.996516 & 0.880745 & 0.141139 & 0.003484 & 0.855377 & 1.468456 & 0.021413 & 0.016776 & 2.696103 & -0.416134 & 3.112237 & 1.000000 & -0.012521 & 1.012521 \\
 29 & 0.996390 & 0.876790 & 0.139010 & 0.003610 & 0.857380 & 1.475529 & 0.021303 & 0.016690 & 2.705795 & -0.413995 & 3.119790 & 1.000000 & -0.012526 & 1.012526 \\
 30 & 0.996265 & 0.872854 & 0.136938 & 0.003735 & 0.859327 & 1.482424 & 0.021193 & 0.016604 & 2.715568 & -0.411865 & 3.127433 & 1.000000 & -0.012530 & 1.012530 \\
 31 & 0.996140 & 0.868939 & 0.134920 & 0.003860 & 0.861220 & 1.489149 & 0.021084 & 0.016518 & 2.725423 & -0.409745 & 3.135167 & 1.000000 & -0.012535 & 1.012535 \\
 32 & 0.996014 & 0.865043 & 0.132953 & 0.003986 & 0.863062 & 1.495712 & 0.020975 & 0.016433 & 2.735360 & -0.407633 & 3.142993 & 1.000000 & -0.012540 & 1.012540 \\
 33 & 0.995889 & 0.861167 & 0.131035 & 0.004111 & 0.864854 & 1.502121 & 0.020867 & 0.016348 & 2.745381 & -0.405530 & 3.150912 & 1.000000 & -0.012544 & 1.012544 \\
 34 & 0.995764 & 0.857311 & 0.129164 & 0.004236 & 0.866600 & 1.508381 & 0.020759 & 0.016264 & 2.755486 & -0.403437 & 3.158922 & 1.000000 & -0.012549 & 1.012549
\end{tabular}
}
\end{ruledtabular}
}
\end{table}
\end{turnpage}

\clearpage

\section{Supplementary Material}

This Supplementary Material shows an example of the sequence of Penrose processes at the decay radius, $\hat r = 1.2$, under the condition it is a turning point of the three particles involved in each process. The calculation considers, at each step, the reduction of the mass and spin of the BH and the consequent increase of the irreducible mass caused by the absorption of the particle ``1'' that falls into the BH. As we have shown, the reduction of the BH spin changes, at each step, the particles' effective potential of radial motion. This effect leads to a minimum, lower limit to the BH spin, under which the boundary conditions of the Penrose process are not fulfilled. 

\begin{figure*}
    \centering
    \includegraphics[width=0.44\hsize,height=20cm]{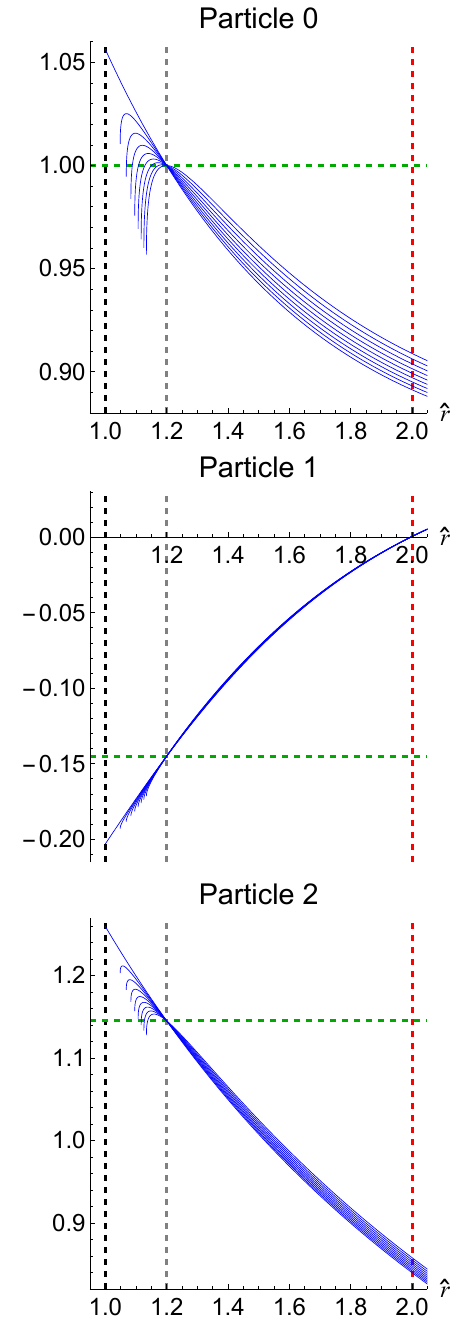}
    \includegraphics[width=0.42\hsize,height=20cm]{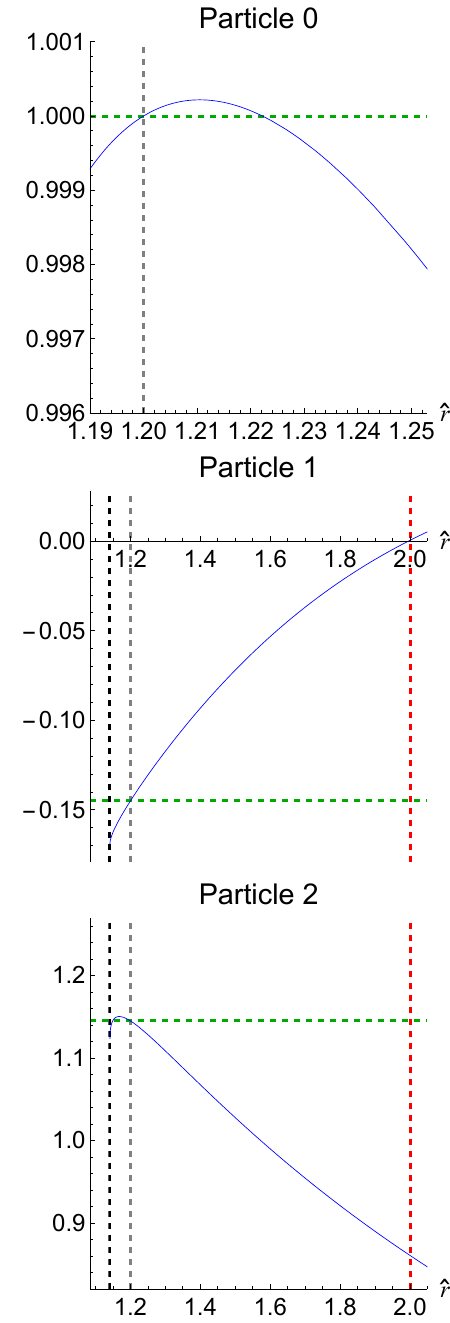}
    \caption{Particle energies, $E_i/\mu_0$ (dashed-green horizontal lines), and effective potentials (blue curves), $V^+_i/\mu_0$, for the sequence of Penrose processes at $\hat r = 1.2$ (see Table II of the paper). Left: the solution of each of the $8$ processes (upper part of Table II of the paper). The difference in particle energies in this part of the sequence is very small, so the eye does not appreciate it in this scale. The sequence ends when particle ``0'' energy corresponds to the effective potential maximum. It occurs at the spin $\hat a = \hat a_{\rm min,0} = 2 \sqrt{\hat r}-\hat r = 0.9909$. Right: solution of the 9th process, if we force the system of equations to give a solution to lower BH spin, $\hat a < \hat a_{\rm min,0}$. It can be seen that the turning point of particle ``0'' lies on the left side of the effective potential.}
    \label{fig:pots}
\end{figure*}

The particle ``0'' is assumed at rest at infinity, so $\hat E_0 = 1$. The other fixed parameters are $\hat p_{\phi 1} = -19.434$, $\mu_2/\mu_1 = 0.78345$, and $\mu_0 = 10^{-2}$. The system of equations consisting of energy and angular momentum conservation at the triple turning point is fulfilled up to the stage when particle ``0'' approaches the top of the effective potential, i.e., when $\hat V^+_{0,\rm max} = \hat E_0 = 1$, as shown in the left panel of Fig. \ref{fig:pots}. This limit for the present decay radius occurs at the BH spin $\hat a_{\rm min,0} = 2 \sqrt{\hat r}-\hat r = 0.9909$, which is reached after $8$ processes (see left panel). The changing BH and particles' parameters at each step are shown in the upper part of Table II of the paper. If we continue the process, the solution of the Penrose process equations leads to particle ``0'' turning point on the left side of the effective potential, which is a non-physical solution since it is assumed to come from infinity. The right panel of the figure shows the solution to the 9th process, which would correspond to $\hat a = 0.9899 < \hat a_{\rm min,0}$.

\end{document}